\documentclass[showpacs, pra,twocolumn,preprintnumbers ,amsmath, amssymb, superscriptaddress, aps]{revtex4-2}
\usepackage{graphicx}
\usepackage{lipsum}
\graphicspath{{Figures/}}
\usepackage{dcolumn}
\usepackage{bm}
\usepackage[colorlinks]{hyperref}
\usepackage{physics}
\usepackage{physconst}
\usepackage{physunits}
\usepackage{subfloat}
\usepackage[export]{adjustbox}

\def\b{\begin{equation}}
	\def\e{\end{equation}}

\begin{document}
	
	
	\title{
		Rashba contribution of 2D Dirac-Weyl fermions: Beyond ordinary quantum regime}
	\author{Ahmed Jellal}
	
	\affiliation{
		Laboratory of Theoretical Physics, Faculty of Sciences, Choua\"ib Doukkali University, PO Box 20, 24000 El Jadida, Morocco
	}
	\affiliation{Canadian Quantum Research Center, 204-3002 32 Ave Vernon,
		BC V1T 2L7, Canada
	}
	
	\author{Dariush Jahani}
	\affiliation{Department of Physics, Sharif University of Technology, P.O. Box 11155-9161, Tehran, Iran	
	}
	
	\author{Omid Akhavan}
	\affiliation{Department of Physics, Sharif University of Technology, P.O. Box 11155-9161, Tehran, Iran		
	}

	\date{\today}
	
	\begin{abstract}
		
		 We study the energy levels of Dirac-Weyl fermions in graphene subject to a magnetic field with Rashba contribution in the minimal length situation. The exact solution for the energy dispersion of Dirac-like charge carriers coupled to the magnetic moments in a (2+1)-dimension is obtained by the use of the momentum space representation. Moreover, as it comes to applications for 2D Dirac-like quasiparticles, we also extend our theory and results in some special cases, showing that the emerging energy spectrum at the high magnetic field limit becomes independent of the Rashba coupling, $\lambda_{R}$, and the band index of Landau levels.
	\end{abstract}
	
	\pacs{03.65.Nk, 03.65.Pm, 04.50.Kd\\
		{\sc KEYWORDS:} Graphene, spin-orbit interaction, minimal length, homogeneous magnetic field, Dirac equation.}
	\maketitle

	\section{Introduction}
	
	The separation of distance scales between UV and IR physics in quantum field theory could no longer exist in high-energy quantum mechanics. For example, in quantum gravity or string theory, UV/IR mixing should be considered, and the separation between them in ordinary quantum mechanics does not hold anymore \cite{1,2,3,4}.
	In other words, under the generalized Heisenberg uncertainty principle (GUP), there is a minimal length {$(\Delta x)_{min}=\hbar\alpha$ \cite{Das2010},} which also contains UV/IR mixing \cite{5,6}. Recently, the minimal length formalism has been the subject of various studies \cite{7,8,9,10,1000,2000,3000,4000}. In this case, it is reasonable to depart from standard quantum mechanics, which considers the relation $[ x_{i},p_{j}]= i\hbar  \delta_{ij}$.

	On the other hand, in low energy quantum mechanics, the Schr\"{o}dinger equation can still be employed in order to consider spin states of the problem, which is important both in high and low energy quantum physics. One specific case is considering the spin states in the presence of relativistic corrections such as Rashba coupling, which itself is derived from the Dirac equation \cite{11,12}. Now, fortunately, the Dirac-like equation in 2D material makes it possible to think of Rashba coupling in the presence of Dirac-like particles within the minimal length formalism. In order to study the solution of a solvable problem in the GUP regime, it is common to convert the associated relations into some standard solvable model in ordinary quantum mechanics \cite{13}.
	
	Among recently mentioned studies about the minimal length concept are some reports on exactly solvable models \cite{16}. In this regard, authors in \cite{17} have obtained the exact solution of the (2+1)-dimensional Dirac equation in the presence of a constant magnetic field. However, the effect of a constant magnetic field is on the orbital motion of electrons, which can be coupled to the spin of the particles \cite{18,19}. Therefore, spin-orbit contribution in the minimal length framework for this problem is worth investigating. Furthermore, the spin-orbit interaction in the presence of an external magnetic field can be described by the Rashba Hamiltonian \cite{20,21,22,23}. The pseudo-spin states of the Dirac-Weyl equation allow one to study the Rashba spin-orbit interaction alongside the Dirac-like particles.
	
	In this letter, we study the corresponding energy dispersion of Dirac fermions in graphene with Rashba spin-orbit interaction within the minimal length formalism. To be more specific, the exact dispersion of fermions in graphene considering the Rashba spin-orbit contribution from a constant magnetic field is obtained. We show that the minimal substitution for introducing the vector potential results in adding a complex effective term to the velocity of light without violating Einstein's theory of special relativity. This complex velocity turns out to affect the energy dispersion relation and might be inferred as the imaginary part of a refractive index showing gain or absorption.
	
	The remaining parts of this letter are organized as follows: In section \ref{Theo}, we consider the theoretical model for the effect of Rashba coupling on the energy dispersion of Dirac-like fermions.  Using the scheme described in \cite{17}, we reduce the problem to a standard one to obtain the final result in section \ref{EnSp}. We discuss potential applications of our model in 2D materials with Dirac-like quasiparticles in section \ref{ReDi}. Finally, we conclude by summarizing the main results.

	\section{Method and theory}\label{Theo}
	
	Despite the fact that numerous types of GUP have been proposed, due to article length constraints, we only investigate the following types of GUP in our current study.   
A new form of GUP based on doubly special relativity (DSR) was proposed in \cite{3000, 2000},  and the algebra generated by the position $x_ i $ and momentum $p_ j $ operators is then presented 
	\begin{align}\label{GUP}
		&	\left[x_{i}, p_{j}\right]=i \hbar\left[\delta_{i j}-\alpha\left(p \delta_{i j}+\frac{p_{i} p_{j}}{p}\right)+\alpha^{2}\left(p^{2} \delta_{i j}+3 p_{i} p_{j}\right)\right]\notag\\
		&	\left[x_{i}, x_{j}\right]=\left[p_{i}, p_{j}\right]=0, \quad i,j=1,2,3
	\end{align}
	where the minimal length is
	\b
	\alpha=\alpha_{0} / M_{\mathrm{Pl} }c=\alpha_{0} \ell_{\mathrm{Pl}} / \hbar
	\e
	$M_{\mathrm{Pl}}$ is the Planck mass, $\ell_{\mathrm{Pl}}=$ $10^{-35} \mathrm{~m}$ is the Planck length, and $M_{\mathrm{Pl}} c^{2} \approx 10^{19}\ \mathrm{GeV}$ is the Planck energy, with $c$ is speed of light and $ \alpha_{0} $ is usually assumed to be less than one,
	and 
	\b
	p^2= p^2_1+p^2_2+ p^2_3.
	\e
	This GUP predicts a maximum visible momenta in addition to the existence of a minimal measurable length and is consistent with string theory, black hole physics, and DSR \cite{733,744,755}. A realization satisfying the algebra \eqref{GUP} can be taken as
\begin{align}
	&
	x_{i}=x_{0 i}\\
	&{p_{i}=p_{0 i}\left(1-\alpha |\bold{p}_{0}|+2 \alpha^{2} |\bold{p}_{0}|^{2}\right)}
	\label{repr}
\end{align}
such that $x_{0 i}$ and $ p_{0 j}$ satisfy the canonical commutation relations 
		\b
	\left[x_{0 i}, p_{0 j}\right]=i \hbar \delta_{i j}
	\e
	 with the momentum
	\b
	p_{0 i}=-i \hbar \frac{\partial } {\partial x_{0 i}}, \quad
{|\bold{p_{0}}|^{2}=p_{0 i} p_{0 i}}	
	\e
	corresponding to $\alpha=0$, i.e., ordinary quantum mechanics.

	Let us apply the above mathematical tools to study the energy of fermions in graphene, which is a hexagonal lattice with two Dirac points that are inequal $K $ and $ K' $ at the zone corners \cite{Novoselov2004,Zhang2006}. Both valleys $ K $ and $ K' $ have linear band crossings in the band structure. 
	The mapping with the spin degree of freedom $s=\{{\uparrow}, {\downarrow}\}$ results in the basis $\psi_{A \uparrow}, \psi_{B \uparrow}, \psi_{A \downarrow}, \psi_{B \downarrow}\}$ for $K$ and $K'$ \cite{1616, 1717}
	\begin{align}
		&\psi_{{A}} \mapsto \psi_{{As}}=\psi_{{A}} \otimes|s\rangle \\
		&\psi_{{B}} \mapsto \psi_{{Bs}}=\psi_{{B}} \otimes|s\rangle.
	\end{align}
	The Hamiltonian for the valley $K$, with Rashba coupling, is given by \cite{Mele2005,MacDonald2006}
	\b\label{100}
	H=v_F \boldsymbol{\sigma} \cdot \mathbf{p}\otimes \mathbb{I}_{s}+\lambda_R(\boldsymbol{\sigma} \times \boldsymbol{s})_{z}
	\e
	where $v_{F}\approx 10^6$ m/s is the Fermi velocity, 
	$\boldsymbol{\sigma}$ and  $\boldsymbol{s}$ are Pauli matrices for the pseudospin (A-B sublattice) and spin degrees of freedom, respectively, $\mathbb{I}_s$ is the $s$-space unit matrix, and $\lambda_R$ represents the Rashba coupling strength. We get the Hamiltonian for the valley $K'$ by substituting  $ \boldsymbol{\sigma} \rightarrow- \boldsymbol{\sigma}^{*}$.
We ignored the intrinsic {spin-orbit coupling (SOC)}, which is approximately $ 12$ $\mu$eV \cite{1616} and very low compared to  $\lambda_R = 0.011$~meV \cite{Mele2005,MacDonald2006}. This depends on the electric field applied perpendicularly to the graphene sheet, which is very strong compared to the intrinsic SOC.

	For our task, we introduce the conjugate momentum 
	\begin{align}
		\boldsymbol{\pi} = \bold{p}+\frac{e}{c}\bold{A}
\label{122}
	\end{align}	
where the vector potential $\bold A$ is chosen in the Landau gauge $  \bold{A}=\left(0,Bx_1\right)$ associated with the magnetic field $\bold B=(0,0,B)$. Based on \eqref{repr}, the components of  $\boldsymbol{\pi}$
are given by
\begin{align}
\pi_{i}=\pi_{0 i}\left(1-\alpha \pi_{0i}+2 \alpha^{2}  \pi_{0i}^{2}\right)	\label{133}
\end{align}	
where $\pi_{0 i}$ represent components of conjugate momentum {$\boldsymbol{\pi}_0=\bold{p}_0 +\frac{e}{c}\bold{A}$} with no deformation ($\alpha = 0$), and $i = 1, 2$. {Now, using \eqref{122} of components \eqref{133} and ignoring the squared term of $\alpha$,  we can rewrite the Hamiltonian \eqref{100} as} 
	\b
	H=
	v_{F}\ {\boldsymbol{\sigma} \cdot \boldsymbol{\pi}_0\left(\mathbb{I}-\alpha \boldsymbol{\sigma} \cdot \boldsymbol{\pi}_0\right)}\otimes \mathbb{I}_{s}
	+
	\frac{\lambda_R}{2}\left(\sigma_1s_2 - \sigma_2s_1\right).	
	\label{defH}
	\e
	It is obvious that changing the Heisenberg algebra will have an impact on all quantum mechanical Hamiltonians. It is clear that when $\alpha = 0$,  \eqref{defH} is reduced to \eqref{100}.
We emphasize that the energy spectrum of massless fermions in graphene subjected to a deformation based on GUP was studied in \cite{17}. It is used to derive an upper bound on the minimal length by comparing it to experimental measurements of Landau levels in graphene.	
	Note that the spin-orbit coupling is used in condensed matter systems to mimic certain expected interactions between charge degrees of freedom and crystalline potentials. In fact, the effective Dirac-Weyl Hamiltonians are not relativistic in the true sense of the word, but they are obtained in a continuum approximation of the lattice dynamics. As a result, the high energy IR/UV limits are constrained by the energetics of the material. In other words, the linear bands for massless particles bend for the real dispersion beyond a certain energy.

	\section{Energy spectrum}\label{EnSp}
	We proceed by transforming the spinor $\phi(x,y)$ associated with \eqref{defH} into
	\b
	\psi(x,y) = \left(\mathbb{I}+\alpha  \boldsymbol{\sigma} \cdot \boldsymbol{\pi}_0\right)\otimes \mathbb{I}_{s}
	\phi(x,y).
	\e
	Then, using the eigenvalue equation 
	$ H \phi(x,y)= E \phi(x,y)$ to the first order of the GUP parameter $ \alpha $ in the basis of the particle and antiparticle wave functions
	 $\phi(x,y)=(\phi_{A\uparrow}, \phi_{B\uparrow}, \phi_{A\downarrow}, \phi_{B\downarrow})^T\ e^{ik_y y}$, $T$ stands for transpose, we obtain
	\begin{widetext}
		\b\label{hamt}
		\begin{pmatrix}
			-E & (v_F-\alpha E)\pi_{0-} & 0 & 0 \\
			(v_F-\alpha E)\pi_{0+}  &-E & -i\lambda_R  &  0 \\
			0 & i\lambda_R & -E & (v_F-\alpha E)\pi_{0-} \\
			0 &   & (v_F-\alpha E)\pi_{0+}  & -E\\
		\end{pmatrix}\phi(x,y)=0
		\e
	\end{widetext} 
where $\pi_{0\pm}=  \pi_{01} \pm i \pi_{02}$ are used.
	We introduce the lowering and raising operators to determine the solutions of the energy spectrum. They are
\begin{align}
		&
	a=\frac{l_{B}}{\sqrt{2} \hbar}\left(\pi_{01}-i \pi_{02}\right)\\ &a^{\dagger}=\frac{l_{B}}{\sqrt{2} \hbar}\left(\pi_{01}+i \pi_{02}\right)
	\end{align}
	satisfying the commutation relation 
	 \b [a,a^{\dagger}]=\mathbb{I}\e
	  and  the magnetic length is $ l_B=\sqrt{\frac{\hbar c}{eB}} $. Then we can map \eqref{hamt} as
	\begin{widetext}
		\b\label{hamt2}
		\begin{pmatrix}
			-E & \hbar \omega_c (v_F-\alpha E) a & 0 & 0 \\
			\hbar \omega_c (v_F-\alpha E) a^\dagger &-E & -i\lambda_R &  0 \\
			0 & i\lambda_R& -E  &   \hbar \omega_c (v_F-\alpha E) a \\
			0 & 0 &  \hbar \omega_c (v_F-\alpha E) a^\dagger & -E\\
		\end{pmatrix}
		\begin{pmatrix}
			\phi_{A\uparrow}\\ \phi_{B\uparrow}\\ \phi_{A\downarrow}\\ \phi_{B\downarrow}
		\end{pmatrix}=0
		\e
	\end{widetext}
	where  $ \omega_c=\frac{\sqrt{2}}{l_B} $ is the cyclotron frequency. This establishes  the set of equations
	\begin{align}
		&-E \phi_{A\uparrow} + \hbar \omega_c (v_F-\alpha E)a \phi_{B\uparrow}=0\\
		& \hbar \omega_c (v_F-\alpha E) a^\dagger \phi_{A\uparrow}
		-E \phi_{B\uparrow} -i\lambda_R\phi_{A\downarrow}=0\\
		& i\lambda_R\phi_{B\uparrow}-E \phi_{A\downarrow}+ \hbar \omega_c (v_F-\alpha E) a\phi_{B\downarrow}=0\\	
		& \hbar \omega_c (v_F-\alpha E) a^\dagger \phi_{A\downarrow}- E \phi_{B\downarrow} =0.
	\end{align}
	Because we have used harmonic oscillator algebra, the basis can be expressed in terms of harmonic states as follows
	\b
	\begin{pmatrix}
		\phi_{A\uparrow}\\ \phi_{B\uparrow}\\ \phi_{A\downarrow}\\ \phi_{B\downarrow}
	\end{pmatrix}=
	\begin{pmatrix}
		\ket{A\uparrow,n-1}\\ \ket{B\uparrow,n}\\ \ket{A\downarrow,n-1}\\ \ket{B\downarrow,n}	
	\end{pmatrix}, \quad n\in \mathbb{N}.
	\e
	As a result, we arrive at the following equation for the energy spectrum
\begin{widetext}
	\b\label{eqeva}
	E^{4} -\left[2n\beta^2(v_F -\alpha E)^2 +   \lambda_{R}^2\right] E^2+n^2\beta^4(v_F -\alpha E)^4
	=0
	\e
		\end{widetext}	where we have set $\beta =\hbar \omega_c$.
	This can be solved to obtain the four band energies listed as below
	\begin{widetext}
		\begin{align}
			&	\label{E11} E_1^s(n,\lambda_{R},\alpha)=\frac{-2n\alpha v_F \beta^2 +\lambda_R + s\sqrt{\lambda_R^2 +4n v_F \beta^2\left( v_F -\alpha \lambda_R\right)}}{2(1-n\alpha^2\beta^2)}\\
			& \label{E22}	E_2^{s'}(n,\lambda_{R},\alpha)=
			\frac{-2n\alpha v_F \beta^2 -\lambda_R + s'\sqrt{\lambda_R^2 +4n v_F \beta^2\left( v_F -\alpha \lambda_R\right)}}{2(1-n\alpha^2\beta^2)}	
		\end{align}
	\end{widetext}
	where different bands are denoted by   $ s,s'=\pm1 $. As we will see, this generalizes a variety of findings from the literature.

	\section{Results and discussions}\label{ReDi}
	
	Now we will take a look at some of the exceptions. To begin, we use the ordinary quantum regime to recover the result for $\alpha=0$ \cite{Jellal2019}
	\begin{align}
		&	E_1^s(n,\lambda_{R})=\frac{\lambda_R + s\sqrt{
				\lambda_R^2 +4n \beta^2 v_F^2}}{2}\\
		&	E_2^{s'}(n,\lambda_{R})=
		\frac{ -\lambda_R + s'\sqrt{\lambda_R^2 +4n \beta^2 v_F^2}}{2}	
	\end{align}
	Second, the Landau levels (LLs) for fermions in graphene can be found by requiring $\alpha=\lambda_R =0$
	\cite{ll1,ll2,ll3,1515}:
	\begin{align}\label{eq22}
		E_{n}^{s}=sv_F{\sqrt{\frac{2e\hbar B}{c} n}
		}.	
	\end{align}
	Additionally, by imposing some restrictions on the magnetic field (high field), we can make more approximations. Thus, in this limit, we obtain
	\b\label{LLs}
	E\approx \frac{v_F}{\alpha}
	\e
	which shows that the dispersion relation is no longer affected by the Rashba coupling or the band index and the magnetic field as well. 
	
	At this level, we have some comments in order. Indeed, according to \eqref{E11} and \eqref{E22}, the evolution of LLs as a function of the magnetic field undergoes a second quantization. It is clear that, in the ordinary quantum limit, the LLs \eqref{eq22} are not equally spaced and the energy level spacing decreases as $n$ increases. To continue, we remark that in the presence of a magnetic field, it may be feasible to identify the coupling strength in the graphene layer using the LLs. The findings could be useful in the development of new graphene-based spintronic devices, especially when external magnetic field, Rashba coupling, and parameter of theory (Planck mass) are used to regulate the electron spin in graphene.

	\begin{figure*}[ht]
		\includegraphics[scale=0.175]{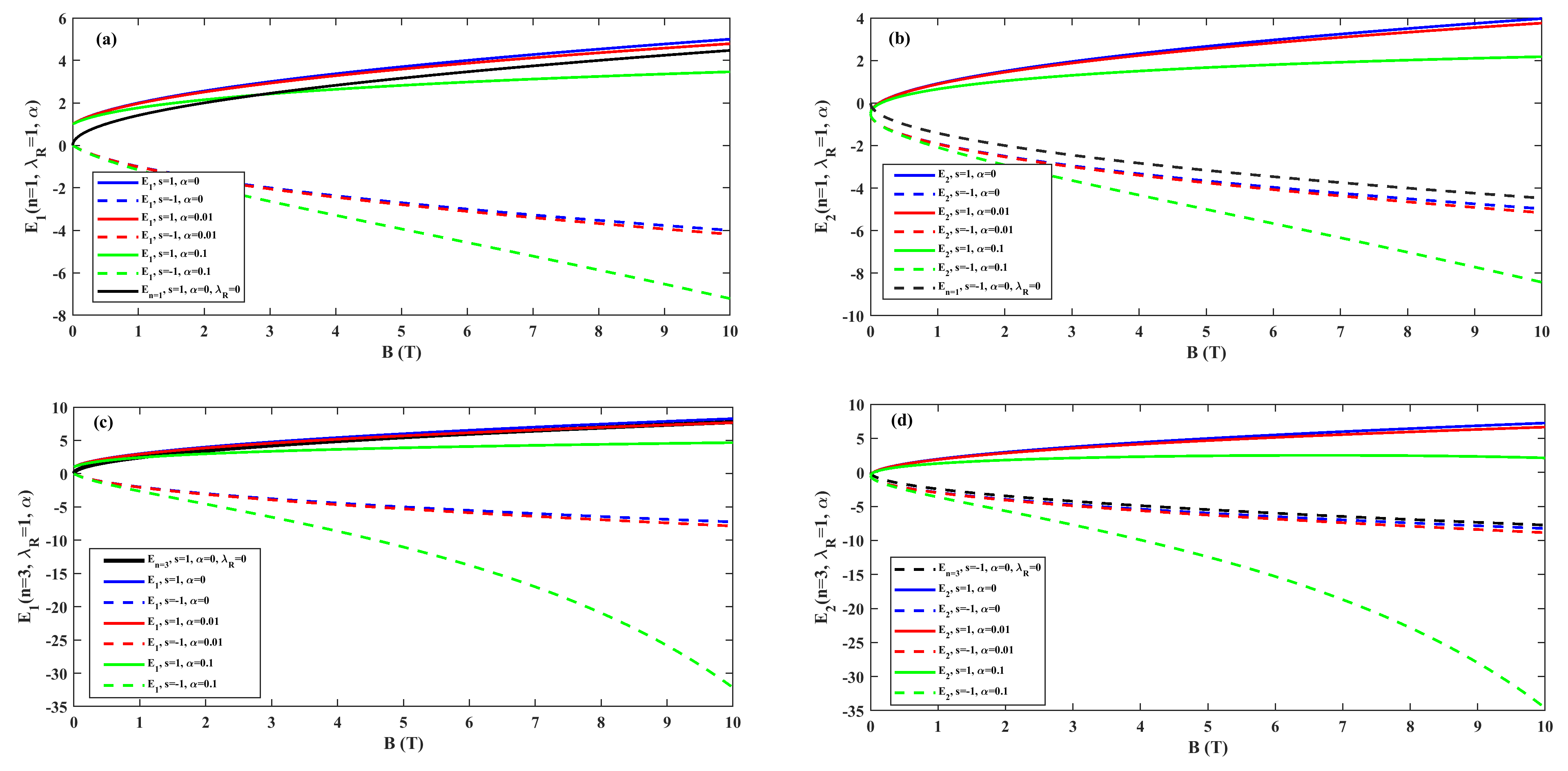}
		\caption{(color online) Energy spectrum  versus the magnetic field $ B $ for two Landau Levels $n=1,3$ and three values of Planck mass $\alpha=0, 0.01, 0.1$ with Rashba coupling $\lambda_{R}=1$.}
		\label{fig1}
	\end{figure*}

	\begin{figure*}[ht]
		\includegraphics[scale=0.18]{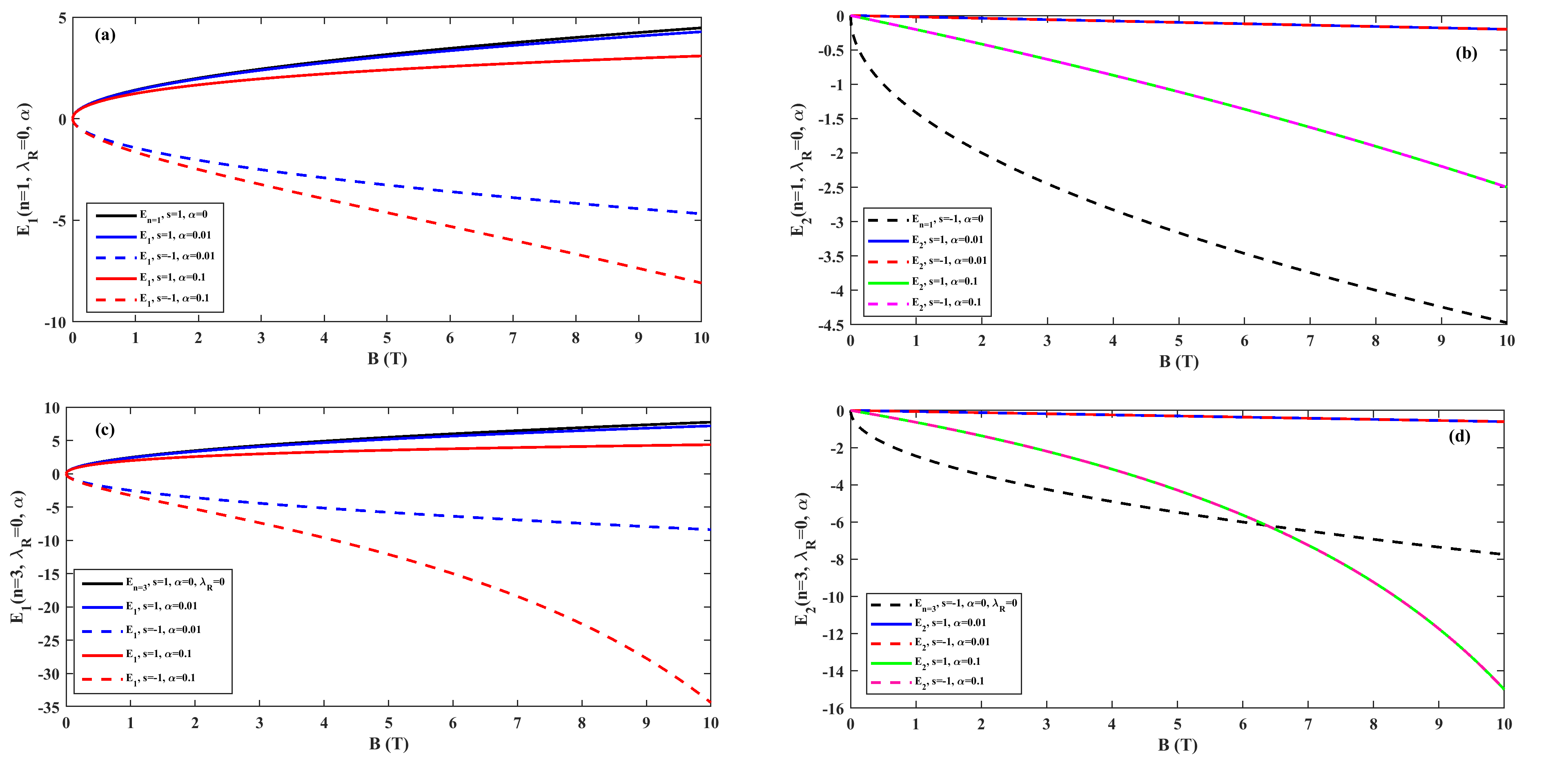}
		\caption{The same as Fig. \ref{fig1} with $\lambda_{R}=0$.}  \label{fig2}
	\end{figure*}

	Fig. \ref{fig1} shows the energy spectrum of Dirac-like charge carriers in graphene as a function of the static magnetic field between $B=1$-$10$ T 
	for the Landau levels  $n=1, 3$  when Planck mass varies as $\alpha=0, 0.01, 0.1$. These results are compared to the conventional Landau levels \eqref{LLs} (black line/dash line).  Figs. \ref{fig1}(a,b) present the four band energies \eqref{E11} and \eqref{E22} when $n=1$, $\lambda=1$ and $\alpha=0$ for $s=+1$ and $s=-1$, respectively. It is apparent  that the energy levels undergo a second quantization or degeneracy for each energy level. For $E_{1}$ when $\alpha$ is increased to $0.01$ in Fig. \ref{fig1}(a), the second-degenerated-band energies separated more which resulted in a little gap between energy levels. This generated gap is more evident when the Planck mass takes a higher amount, $\alpha=0.1$. In the case of $E_{2}$, the same results are obtained unless for a higher amount of $\alpha$ no gap is reported, as shown in  Fig. \ref{fig1}(b). Furthermore, when $n=3$ is considered, the second-degenerated-energy bands  $E_{1}$ and $E_{2}$ get more energy in comparison with $n=1$, but no gap between energy bands is obtained, see Figs. \ref{fig1} (c,d).
	
	Now, we will look at the calculation for $\lambda_R=0$ when $\alpha=0, 0.01, 0.1$ in the magnetic field interval shown in Fig. \ref{fig2}. It is worthy of mentioning that the case of $\lambda_R=0$ and $ \alpha = 0 $ represents the conventional Landau levels, which have been reported, for example, in \cite{ll1,ll2,ll3,1515}. The band energies are degenerated for the second time for $E_1 $ when $n=1$ (Fig. \ref{fig2}a) and $n=3$ (Fig. \ref{fig2}c). However, no second band energy degeneracy can be seen for $E_ 2 $ (Figs. \ref{fig2} (b,d). Furthermore, by increasing $\alpha$, the $E_1$ degenerated-energy bands have more energy than the $E_2$ energy levels because $E_2$  do not undergo second quantization.

	\section{CONCLUSION}
	In summary, using the momentum space representation, we obtained the explicit solution for the energy dispersion of Dirac-like particles in graphene under the generalized Heisenberg uncertainty principle and Rashba spin-orbit interaction. We showed that our results for the dispersion relation recover Landau energy levels using the ordinary quantum regime for $\alpha=\lambda_R=0$. Then, it was shown that in a strong magnetic field, the energy spectrum comes out to be independent of the Rashba coupling but is still a function of the minimal length. This Rashba-independence property of Dirac fermions may be particularly interesting for Klein tunneling of Dirac electrons through magnetic barriers.

	Moreover, the general energy dispersion relation shows that the band velocity is modified and can be examined, for example, by infrared spectroscopy. In the end, we should note that one recent work that reported on the band velocity of Landau levels in graphene shows a measured band velocity of around $1.1\times 10^{6}$ m/s \cite{25}. It is argued in this paper that the lack of precise scaling is due to many-particle contributions to the infrared transitions. Therefore, spin orbit coupling could be considered to have a measured effect on the Fermi velocity of Dirac-like particles in graphene among many-body effects in this regard.
	
\section*{Author contributions}
	All authors have contributed equally to the paper.

\section*{Data Availability Statement} 
This manuscript has no
	associated data or the data will not be deposited. [Authors’
	comment: The data that support the findings of this study
	are available on request from the corresponding author].
	
	\newpage
	\bibliographystyle{plain}

\begin{thebibliography}{a}
		\addcontentsline{toc}{chapter}{Bibliographie}
		\bibitem{1} Gianluca Mandanici,  Wave propagation and IR/UV mixing in noncommutative spacetimes,  hep-th/0312328 (2003).
		\bibitem{2} S. Vaidya and B. Ydri,
		Nucl.
		Phys. B 671, 401 (2003).
		\bibitem{3} Harald Grosse,  Harold Steinacker, and Michael Wohlgenannt,
		JHEP 04,  023 (2008).
		
		\bibitem{4} Esperanza L\'opez,
		JHEP 09, 033 (2003).
		
\bibitem{Das2010} Saurya Das, Elias C. Vagenas, and Ahmed Farag Ali, Phys. Lett. B 690. 407 (2010).
		\bibitem{5}  Abdel Nasser Tawfik and Abdel Magied Diab,
		Rep. Prog. Phys. 78, 126001 (2015).
		\bibitem{6}  Michele Maggiore,
		Phys. Lett. B 304, 65 (1993).
		\bibitem{7} Kh. Nouicer,
		J. Phys. A: Math.  Gen. 38, 10027 (2005).
		\bibitem{8}Omid Akhavan,
		Acta Scientific Applied Physics 2, 34 (2022).
		\bibitem{9} Achim Kempf, Gianpiero Mangano, and Robert B. Mann, 
		Phys. Rev. D 52, 1108 (1995).
		\bibitem{10}T. V. Fityo,  I. O. Vakarchuk, and V. M. Tkachuk, 
		J. Phys. A: Math.  Gen.  39, 2143 (2006).

			\bibitem{1000} Saurya Das and Elias C. Vagenas,  Phys. Rev. Lett. 101, 221301 (2008).
			
	\bibitem{3000} Ahmed Farag Ali, Saurya Das, and Elias C. Vagenas, Phys. Lett. B 678, 497 (2009).
					
		\bibitem{4000} Saurya Das, Elias C. Vagenas, and Ahmed Farag Ali, Phys. Lett. B 690, 407 (2010).
				
		\bibitem{2000} Ahmed Farag Ali, Saurya Das, and Elias C. Vagenas, Phys. Rev. D 84, 044013 (2011).
		

		
		\bibitem{11} A. Manchon, H. C. Koo, J. Nitta, S. M. Frolov, and  R. A. Duine,
		Nature Mater. 14, 871 (2015).
		
		\bibitem{12}Charles Tahan and Robert Joynt,
		Phys. Rev. B 71, 075315 (2005).
		\bibitem{13}  G. L\'evai,
		J. Phys. A: Math.  Gen. 39, 10161 (2006).
		
		\bibitem{16} 
		M. F. Gusson, A. Oakes O. Gonçalves, R. O. Francisco, R. G. Furtado, J. C. Fabris, and J. A. Nogueira,
		Eur. Phys. J. C 78, 179 (2018).
		\bibitem{17} L. Menculini, O. Panella, and P. Roy,
		Phys. Rev. D 87,  065017 (2013).
		\bibitem{18} P. Kurian and C. Verzegnassi,
		Phys. Lett. A 380, 394 (2016).
		\bibitem{19}Yu G. Semenov and K. W. Kim,
		Phys. Rev. B 67, 073301 (2003).
		\bibitem{20}Laurens W. Molenkamp,  Georg Schmidt, and Gerrit E. W.  Bauer,
		Phys. Rev. B 64, 121202(R)   (2001).
		\bibitem{21}Daniel L. Campbell, Gediminas Juzeliunas, and Ian B. Spielman,
		Phys. Rev. A 84, 025602  (2011).
		\bibitem{22} 
		A. D. Caviglia, M. Gabay, S. Gariglio, N. Reyren, C. Cancellieri, and J.-M. Triscone,
		Phys. Rev. Lett.  104, 126803  (2010).
		\bibitem{23}Qing-feng Sun,  Jian Wang, and Hong Guo,
		Phys. Rev. B 71,  165310  (2005).
		\bibitem{25} 
		Z. Jiang, E. A. Henriksen, L. C. Tung, Y.-J. Wang, M. E. Schwartz, M. Y. Han, P. Kim, and H. L. Stormer,
		Phys. Rev. Lett. 98, 197403 (2007).
		
		\bibitem{733}
		M. S. Ma and R. Zhao, J. Math. Phys. 55, 082109 (2014).
		\bibitem{744}
		P. Bargueño and E. C. Vagenas, Phys. Lett. B 742, 15 (2015).
		\bibitem{755}
		A. N. Tawfik and E. A. El Dahab, Int. J. Mod. Phys. A 30, 1550030 (2015).
		\bibitem{Novoselov2004}		
		K. S. Novoselov, A. K. Geim, S. V. Morozov, D. Jiang, Y. Zhang, S. V. Dubonos,  I. V. Grigorieva,
		and A. A. Firsov, Science 306, 666 (2004).
		\bibitem{Zhang2006}	Y. Zhang, Y.-W. Tan, H.L. Störmer, P. Kim, Nature 438,
		201 (2005).
		\bibitem{1616}   S. Konschuh,  M. Gmitra,  and  J. Fabian,   Phys. Rev. B 82,
		245412 (2010).
		\bibitem{1717}  Z. Qiao,  H. Jiang,  X.  Li,  Y. Yao,  and  Q. Niu,
		Phys. Rev. B 85,
		115439 (2012).
		
		\bibitem{Mele2005}	
		C. L.	 Kane  and  E. J. Mele,  Phys. Rev. Lett. 95, 226801 (2005).
		
		\bibitem{MacDonald2006} H.	Min, J. E. Hill, N. A. Sinitsyn, B. R. Sahu, L. Kleinman L,  and
		A. H.	MacDonald,    Phys. Rev. B 74, 165310 (2006).
		
		\bibitem{Jellal2019} 	R. Houça and A. Jellal,  Phys. Scr. 94, 105707  (2019).
		\bibitem{ll1} D. L. Miller, K. D. Kubista, G. M. Rutter, M. Ruan, W. A. de Heer, P. N. First, and J. A. Stroscio, 
		Science 324(5929), 924 (2009). 
		\bibitem{ll2} Y. J. Song, A. F. Otte, Y. Kuk, Y. Hu, D. B. Torrance, P. N. First, W. A. de Heer, H. Min, S. Adam, M. D. Stiles, A. H. MacDonald,
		and J. A. Stroscio, 
		Nature 467(7312), 185 (2010).
		\bibitem{ll3} G. Li, A. Luican, and E. Y. Andrei, 
		Phys. Rev. Lett. 102(17), 176804
		(2009).
		
		\bibitem{1515} A. H. Castro Neto,  F. Guinea, N. M. R. Peres, K. S. Novoselov,  and A. K.
		Geim,    Rev. Mod. Phys. 81,  109 (2009).
		
		
		%
		%
		
		
	\end{thebibliography}
	
\end{document}